# Revealing the tribological stress field by using deformation twins as probes

Antje Dollmann[1,2], Alexander Dyck[3], Claudius Klein[3], Alexander Kauffmann[1], Thomas Böhlke[3], Christian Greiner[1,2]

[1] Institute for Applied Materials (IAM), Karlsruhe Institute of Technology (KIT), Kaiserstraße 12, 76131 Karlsruhe, Germany

[2] IAM-ZM MicroTribology Center (µTC), Straße am Forum 5, 76131 Karlsruhe, Germany

[3] Continuum Mechanics, Institute of Engineering Mechanics (ITM), Karlsruhe Institute of Technology (KIT), Kaiserstraße 12, 76131 Karlsruhe, Germany

## Abstract

Microstructural evolution in metallic materials feedbacks with the loading conditions and influences the life time of parts and components. Therefore, the deformation mechanisms have to be fundamentally understood. Tribological loading causes a non-trivial, position-dependent, moving stress field. We present a systematic study on the influence of the complexity of the implemented material models on the calculated stress field. For the stress field validation, results of tribological experiments on single crystals with the activation of deformation twins are used. The resolved shear stresses calculated with the stress field models have to be highest on the experimentally identified twin systems. From this combination of modelling and experiment, it clearly follows that a stress field model considering plasticity is required. The widely used Hamilton stress field for tribological loading is limited due to only considering elastic strains. Here, the predictive quality of the stress field is sensitive to the assumed yield strength, work hardening and plastic anisotropy. Certain stress field models are close to the experimental data, but none completely replicate them. These results highlight that the model type and parameters have to be carefully determined in order to be able to predict how a metallic material deforms due to a sliding load.

## Main body

### Introduction

Tribological contacts are ubiquitous. It is virtually impossible to find a technical application without any relation to tribology. Unfortunately, tribological phenomena are complex and have an abundance of influencing parameters. To make matters worse, tribological properties are not material constants alone, but system properties making it necessary to consider the entire tribological system. Among the many parameters influencing tribological contacts, we here focus on the subsurface deformation layer developing in materials with plastic behaviour. Argibay et al. illustrated a friction feedback loop [1]. This concept states that the friction coefficient influences the stress field resulting in microstructural evolution which itself feedbacks with the friction coefficient [1,2]. While this demonstrates the complexity and interwovenness of tribological processes, it also means that a classical microstructure-properties relation exists. However, the microstructure does undergo drastic changes as the tribological load is applied. Such microstructural evolution is not only known to change friction forces, but also to be the origin of cracks and therefore wear particle formation [3–5] resulting in the failure of engineering components.

In general, the microstructural evolution in metallic materials is caused by plastic deformation mechanisms, i.e., predominately dislocation slip and deformation twinning. These fundamental mechanisms are activated by (resolved) shear stresses (RSS) reaching a material-specific, critical value $\tau_c$. Classically, $\tau_c$ is experimentally determined by quasistatic, monotonic, uniaxial mechanical tests. When tribological systems are considered, inhomogeneous, multi-axial, position-dependent stresses, which are moving with the counter body have to be considered. Thus, analysing the activation and evolution of deformation mechanisms requires an adequate description of the stress field.

As this matter has been recognized for several decades, it is no surprise that several approaches exist in the literature. In the following, the most common stress field models are briefly summarized and arranged in order of their complexity. Literature does not address the limits or the range of usability of these models. This is a challenging task and desperately needed. Also, here we are only able to give a broad stroke summary.

The most straight forward approach to the stress acting under a tribological load is to only consider singular components of the stress tensor. Researchers have relied on a pure shear stress or a pure normal stress in normal or sliding direction [6–8]. This highly simplifies the calculation of the active slip system while at the same time neglecting the lateral stress distribution, multiaxiality and any type of plastic material behaviour.

The next level of complexity is a multiaxial analytical, static solution. Hamilton's approach superimposes a Hertzian contact with a shear stress proportional to the friction coefficient assuming isotropic, linear-elastic material behaviour [9]. Despite its limitations, this stress field model is often used in the literature to rationalize microstructural evolution in metallic materials [10–15]. The Hamilton solution was also employed to investigate dislocation-mediated plastic deformation using discrete dislocation dynamics simulations [16]. Nevertheless, prior work has also demonstrated limitations of the Hamilton solution [17].

As metallic materials undergo plastic deformation, models assuming purely linear-elastic behaviour largely overestimate the stresses acting in the contact and their penetration depth as both depend on plastic flow. The influence of plastic deformation was studied by various simulation methods on very different length scales. These non-static frameworks have the advantage of a moving counter body, capturing the history of the contact. For example, a correlation between the deformation mechanism(s) and the stress field was investigated by molecular dynamic simulations (MD) [18]. The computational costs for MD restrict the simulations to a few million atoms and very short time periods. The high strain rates applied in MD favour deformation mechanisms like stacking fault formation and deformation twinning [19].

In order to tackle larger sample sizes, finite element simulations (FEM) were used. Literature using classical FEM calculating the stress field is limited [20,21]. In recent studies, crystal plasticity FEM simulations (CP-FEM) was used to simulate the plastic response of a material subjected to tribological loading [22–25]. The simulative results were compared to experiments regarding wear track depth, material pile-up and crystal rotation [22–24,26]. While for these parameters a reasonable agreement between simulation and experiment was reported, when considering the materials' microstructures, crystallographic rotations predictions in terms of their sign were possible, but the absolute size of the affected subsurface regions and its values were off.

For all of the above as well as for all future simulations, there is one constant challenge: How to validate the simulative stress field predictions experimentally. Our approach follows two prongs. First, we assume that a useful stress field model has to be able to accurately predict the resolved shear stresses on twin systems. Second, we investigate how considering different material parameters change the stress field calculated with a finite element-based method. We then combine both prongs and calculate the resolved shear stresses on all possible twin systems. This has to agree with experimentally identified twin systems, thereby validating the stress fields usefulness.

Results and discussion

The stress field under tribological load is little understood. Therefore, material parameters are tested and their influence on the stress field evaluated. The material parameters considered are yield strength, work hardening behaviour and elastic and plastic anisotropy. Details concerning the FE simulations can be found in the Supporting Information (SI). The chosen material parameters correspond to experimentally determined values for CoCrFeMnNi, as this was the alloy employed for our reference experiments due to its ideal stacking fault energy and propensity for deformation twinning. Isotropic material models with different yield strength were investigated – 125 MPa labelled as LY (low yield strength) and 360 MPa labelled as HY (high yield strength) [27]. In addition, three different work hardening behaviours are considered. These are: no work hardening (noW), low work hardening (LW) [28] and high work hardening (HW) [28]. The elastic and plastic anisotropy was investigated using crystal plasticity (CP) model (labelled CP-FEM). Here, no work hardening was considered and a critical resolved shear stress for dislocation motion of 43 MPa was used [29].

The von Mises stresses for the isotropic material models are presented in Figure 1, demonstrating the effects of different material parameters. For the sake of clarity, the von Mises stresses calculated with the material model CP-FEM are presented in the Supporting Information.

Comparing the panels shown in in Figure 1, one notices one communality: The lowest z-position of the 100 MPa isoline always is at a depth of around 100 µm. This suggests that the depth of the stress field is marginally influenced by the material model. It has to be mentioned that the depth of the stress field is orders of magnitude deeper than the observed microstructural evolution [17]. Reasons for this tremendous discrepancy between calculated stress fields and experimental reality might be found in the fact that not all dislocation activity was or possibly can be measured; or that for example surface roughness might influence the depth of the stress field much more than one would initially assume. Significant deviations between the different material models are seen in the position and the value of the maximum of the von Mises stresses. The Hamilton model yields a maximum von Mises stress of 452 MPa in a depth of 17 µm. The HYLW model results in a maximum of 362 MPa at a depth of 15 µm. In contrast, considering a lower yield strength (LYLW) results in a maximum of 181 MPa at the surface. When considering different work hardening behaviour for the low yield stress case, the following is observed: high work hardening results into a maximum of 197 MPa in a depth of 5 µm, whereas the maximum for no work hardening is 125 MPa and at the surface. A maximum von Mises stress at the surface agrees with the experimental data as the most pronounced microstructural evolution is observed in the near surface region. In literature, a multiaxiality factor was introduced in order to describe the gaping discrepancy between the von Mises stress maximum according to the Hamilton model and the pronounced microstructural changes directly beneath the surface [12]. While this approach was successful, it seems much more likely to us that plasticity must be considered than trying to tweak a purely elastic stress field.

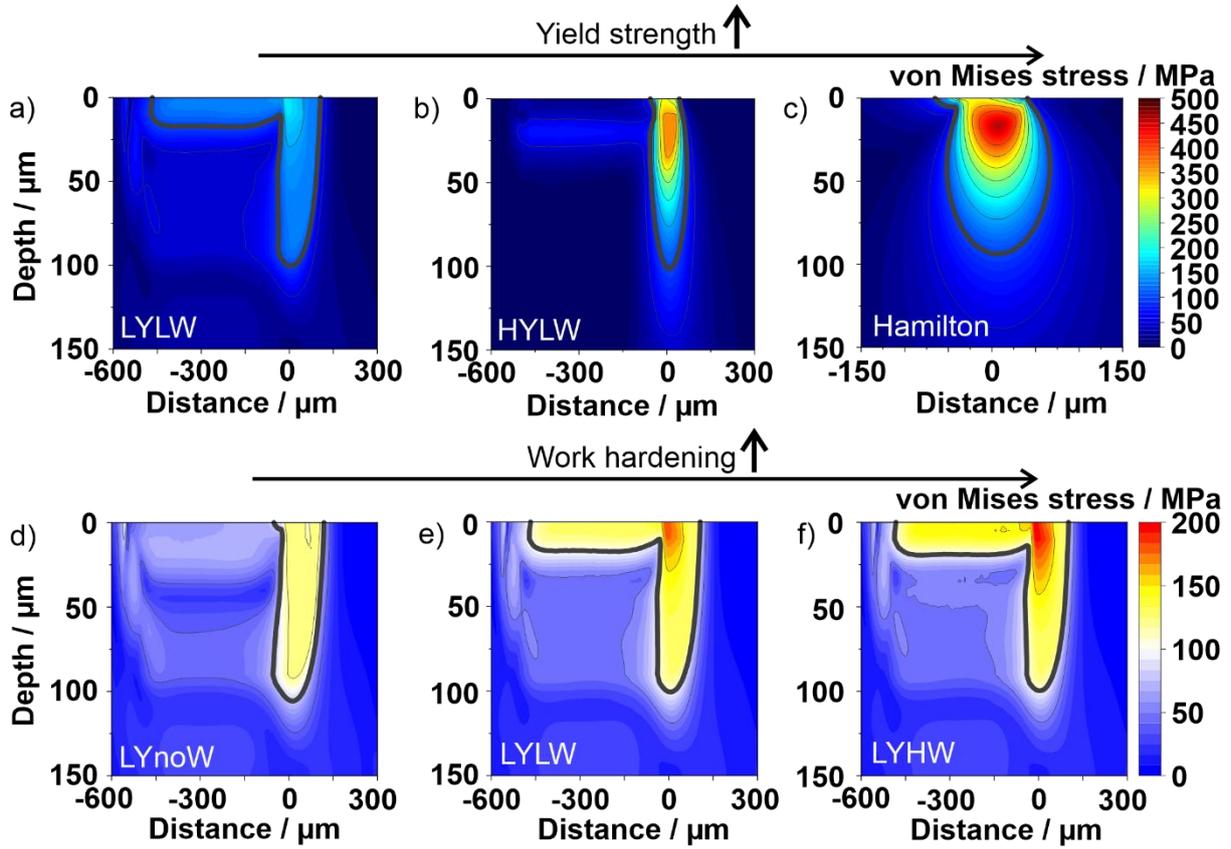

Figure 1. Von Mises stresses for the isotropic material models. The first row shows the influence of increasing yield strength from left to right. The von Mises stress using the material model LYLW is shown in a), HYLW in b) and Hamilton in c). Increasing work hardening is investigated in the second row form left to right. Hereby, the yield strength was kept constant. The material model LYnoW is investigated in d), LYLW in e) and LYHW in f). Different color-coding was used for a)-c) and d)-f) to show the different scaling. The abbreviations HY and LY stand for high yield strength and low yield strength, respectively. noW, LW and HW indicate the degree of work hardening: no work hardening, low work hardening and high work hardening. The spherical counter body starts sliding at the distance of -500 μm expect for the Hamilton model, where the indenter remains at 0 μm. The current loading position is at 0 μm. The friction coefficient is 0.2.

As every stress field model should to be able to successfully predict experimental reality, we here approach such validation by calculating the resolved shear stresses for experimentally observed deformation twinning systems. Deformation twins are chosen as probes as they are indicative of the plane and direction they were formed on and can be analyzed by ex-situ microstructural investigations. This is in contrast to dislocation motion where for example cross-slip, annihilation and bidirectionality leaves some ambiguity. We have earlier described single trace tribological experiments on face-centred cubic CoCrFeMnNi single crystals with defined normal (ND) and sliding directions (SD) [17]. In two out of five crystallographic orientations investigated, deformation twins were observed within the deformation layer. These results are summarized in *Table 1*.

Table 1. Summary of the occurrence of deformation twins in single crystalline CoCrFeMnNi under tribological loading. The determined twin systems are mentioned. The corresponding STEM and HR-TEM images can be found in [17].

|  | ND[00$\bar{1}$]/ SD[$\bar{1}$10] | ND[0$\bar{1}\bar{1}$]/ SD[100] | ND[0$\bar{1}\bar{1}$]/ SD[0$\bar{1}$1] | ND[0$\bar{1}\bar{1}$]/ SD[$\bar{2}\bar{1}$1] | ND[00$\bar{1}$]/ SD[100] |
|---|---|---|---|---|---|
| Twins | ✓ | ✓ | ✗ | ✗ | ✗ |
| Twin system | ($\bar{1}$11)[$\bar{1}$1$\bar{2}$] | ($\bar{1}\bar{1}\bar{1}$)[2$\bar{1}\bar{1}$] | - | - | - |

The results for the maximum resolved shear stresses as calculated from the five stress fields models - Hamilton, HYLW, LYnoW, LYLW, LYLW, LYHW and CP-FEM – are represented in Figure 2. The Hamilton model is given as a reference and as it is the most commonly taken in literature, albeit these results were already presented in [17]. In this plot, each bar represents the maximum RSS among all twin systems for one specific crystallographic orientation and material model, without providing information about the specific twin system. The RSS values on the experimentally, identified twin systems are represented by the green and blue lines for the RSS for ($\bar{1}$11)[$\bar{1}$1$\bar{2}$] and ND[00$\bar{1}$]/SD[$\bar{1}$10] ($\tau_{(\bar{1}11)[\bar{1}1\bar{2}]}$) as well as for ($\bar{1}\bar{1}\bar{1}$)[2$\bar{1}\bar{1}$] and ND[0$\bar{1}\bar{1}$]/SD[100] ($\tau_{(\bar{1}\bar{1}\bar{1})[2\bar{1}\bar{1}]}$), respectively.

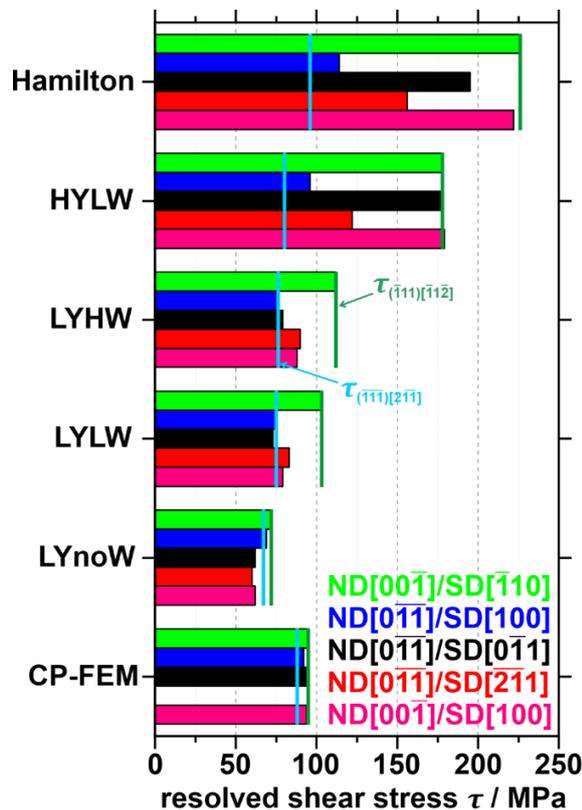

Figure 2. Maximum resolved shear stresses for the experimentally investigated crystal orientations and six investigated material models: Hamilton, HYLW, LYHW, LYLW, LYnoW and CP-FEM. The red bar for CP-FEM is missing based on lacking crystallographic symmetry (see SI). The experimentally identified twin system for the crystal orientation ND[00$\bar{1}$]/SD[$\bar{1}$10] is ($\bar{1}$11)[11$\bar{2}$]. ($\bar{1}\bar{1}\bar{1}$)[2$\bar{1}\bar{1}$] is the experimentally identified twin system for the crystal orientation ND[0$\bar{1}\bar{1}$]/SD[100]. The resolved shear stresses on these two twin systems ($\tau_{(\bar{1}11)[\bar{1}1\bar{2}]}$ and $\tau_{(\bar{1}\bar{1}\bar{1})[2\bar{1}\bar{1}]}$) are marked by the green and blue lines for each material model.

First, the results based on the Hamilton model are summarized to serve as a reference. The focus hereby is on the experimentally identified twin systems. The twin system $(\bar{1}11)[\bar{1}1\bar{2}]$ for ND[00$\bar{1}$]/SD[$\bar{1}$10] exhibits the highest overall RSS. In contrast, $\tau_{(\bar{1}11)[2\bar{1}\bar{1}]}$ for ND[0$\bar{1}\bar{1}$]/SD[100] is not the highest RSS for this orientation and it is lower than all the other maximum RSS of the investigated crystallographic orientations. Although, the Hamilton model shows the correct result for ND[00$\bar{1}$]/SD[$\bar{1}$10], it does not provide the correct results for ND[0$\bar{1}\bar{1}$]/SD[100], as other twin systems, which were experimentally not activated, have higher RSS. Therefore, the Hamilton model is invalid for the given tribological setting when assuming a single, material specific critical activation stress $\tau_c$.

The effect of yield strength is investigated considering the material models: Hamilton, HYLW and LYLW. As expected, a decrease in yield strength results in lower resolved shear stresses, which is dependent on the crystallographic orientation. Comparing Hamilton and LYLW, the highest decrease in RSS of 60 % is calculated for ND[00$\bar{1}$]/SD[100] and the smallest of 22 % for ND[00$\bar{1}$]/SD[$\bar{1}$10]. Considering different work hardening behaviors, while keeping LY constant, show qualitatively similar effects as the yield strength variation. Comparing noW and HW shows the greatest difference of 40 MPa (35%) for ND[00$\bar{1}$]/SD[$\bar{1}$10] and the lowest of 7 MPa (10%) for ND[0$\bar{1}\bar{1}$]/SD[100]. As demonstrated in [17], the position of the resolved shear stress maximum with respect to the indenter position varies with the crystallographic orientation and selected twin system. This variation in position has an influence on the reported differences in resolved shear stress.

For investigating the effect of elastic and plastic anisotropy, no work hardening was considered in CP-FEM to ensure comparability with LYnoW and to avoid other influencing effects than the anisotropy. To strengthen that, the critical RSS for dislocation motion within the CP-FEM framework is equal to LY divided by the Taylor factor. Only four out of the five crystallographic orientations were simulated, see SI. In comparison to LYnoW, the RSS using CP-FEM is increased by at least 31 %. The CP-FEM model could not reflect the experiments as crystallographic orientations without experimentally verified twinning have higher RSS than $\tau_{(\bar{1}11)[2\bar{1}\bar{1}]}$ and furthermore, the $\tau_{(\bar{1}11)[2\bar{1}\bar{1}]}$ is not the highest value within ND[0$\bar{1}\bar{1}$]/SD[100]. Most likely, incorporating work hardening to the CP-FEM model would be also a promising approach as this material model most closely reflect reality. However, choosing the proper microscopic strain hardening model and parameters is a challenge. Furthermore, the stress release and concentration by deformation twinning is then still not mapped and might still impact on the results. As shown in [31], even with a symmetric crystallographic orientation and simple indentation, reproducing the experiment using CP-FEM is not straight forward. Therefore, the isotropic material models LYnoW and LYLW show the best results with the given settings, specifically considering the substantially increased complexity of CP-FEM.

Finally, the critical shear stress for twinning in CoCrFeMnNi at RT has to be discussed. The lowest experimentally measured value is 110 MPa [30]. $\tau_{(\bar{1}11)[2\bar{1}\bar{1}]}$ calculated with all stress field models is lower than 110 MPa. The difference could arise through a stress concentration not covered by the simulation or a possibly limited comparability between twin formation under uniaxial and multiaxial stress.

Besides the RSS approach, the experimentally obtained wear track width has been compared to the widths of the plastic strain on the surface in the transversal direction in the simulations (Figure 3). In contrast to experimental strain field mapping, measuring the wear track width is easily accessible and only requires optical microscopy. Hamilton and HYLW are missing as these did not result in any plastic strain or plastic strain on the surface. Work hardening changes mainly the values of the maximum plastic strain and not the width as shown in Figure 3a. The full width half maximum (FWHM) lies for all three material models calculated with LY between 100 and 110 µm (finite element size of 5 µm). The mean experimental wear track widths vary between 97 and 104 µm, therefore, the simulations with LY is closest to the experimental values. Considering the width of the accumulated slip for CP-FEM in Figure 3b shows a perfect match with the wear track width. From this perspective, CP-FEM is in best agreement with the experimentally measured wear track width. The approach of comparing the experimental wear track width with the extent of equivalent strain or accumulated slip yielded results consistent with the RSS discussion above. It did, however, not allow to further discard material models.

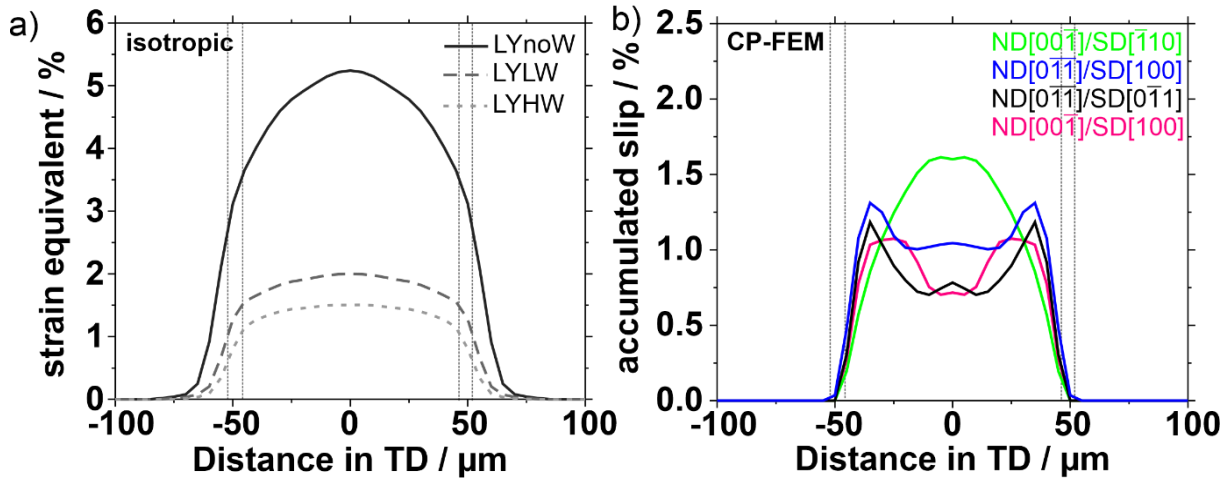

Figure 3. Equivalent strain/accumulated slip on the surface perpendicular to the sliding direction (=transverse direction (TD)). The range of the wear track width from the experimental data is represented by the dotted vertical lines. a) influence of the investigated work hardening behaviours with low yield strength (125 MPa); b) influence of the crystal orientation dependency on the accumulated slip (CP-FEM model).

Conclusion

In summary, six material models were applied to simulate the stress field under tribological loading, five of which were using the finite element method. The aim was to make sense of the microstructural evolution observed in metals and alloys. Yield strength, work hardening and elastic and plastic anisotropy were varied to investigate their respective effects. Considering the von Mises stress, it was observed that the von Mises stress maximum decreases with increasing proportion of plastic deformation and shifts towards the surface; in certain cases, it reaches the surface. The Hertzian model's prediction of a subsurface stress maximum several micrometers beneath the surface is highly inaccurate in the context of plastic deformation. To analyze which model is most suitable (at minimum effort in model and parameter selection) to predict the microstructural evolution under tribological loading, two different methods were used: i) A comparison between the experimentally measured wear track width and the width of the strain equivalent/accumulated slip within the simulations; ii) Can the stress field models predict the formation of the experimentally observed twins. The first method is experimentally faster and allowed to exclude the stress field models, with no or little plasticity. All other material models agreed with the experimental results. The second method is significantly more time-consuming. At the same time, twins are very reliable probes for the tribological stress field. None of the investigated stress field models was able to perfectly predict the experimentally observed twins. This limitation may be partly due to the use of a rather simplified FEM set-up. The resolved shear stress decreases with increasing plasticity, as expected. However, the extend of this decrease is dependent on the crystallographic orientation – also for the isotropic material models - and the position of the resolved shear stress maximum in respect to the indenter. Albeit not being perfect, the results clearly show the great improvement possible for calculating the resolved shear stresses when using an isotropic material model compared to the classical Hamilton approach.

## Methods

The experimental data used for validation were published before and the reader is referred to [17] for all experimental details.

In the following, the relevant equations for solving the boundary value problem of a sample under tribological loading using FEM are summarized. The discretized equations are solved by the commercial FEM program Abaqus [32]. The balance equation governing the evolution of stresses and strains within the sample and the counter body is the balance of linear momentum, where inertia terms due to the motion of the indenter and volume forces are neglected. Due to the balance of angular momenutm, the stress tensor $\boldsymbol{\sigma}$ is symmetric, i.e., $\boldsymbol{\sigma} = \boldsymbol{\sigma}^\mathrm{T}$.

A small strain approximation is used, where the strain tensor is denoted by $\boldsymbol{\varepsilon}$. To incorporate plasticity, the strain tensor is additively decomposed into elastic strains $\boldsymbol{\varepsilon}_\mathrm{e}$ and plastic strains $\boldsymbol{\varepsilon}_\mathrm{p}$ as [33]

$$\boldsymbol{\varepsilon} = \boldsymbol{\varepsilon}_\mathrm{e} + \boldsymbol{\varepsilon}_\mathrm{p}. \qquad \text{Equation 1}$$

The Cauchy stresses $\boldsymbol{\sigma}$ are computed using Hooke's law, i.e., a linear relation between elastic strains $\boldsymbol{\varepsilon}_\mathrm{e}$ and the stresses by means of the stiffness tensor $\mathbb{C}$ is used, which reads [33]

$$\boldsymbol{\sigma} = \mathbb{C}[\boldsymbol{\varepsilon} - \boldsymbol{\varepsilon}_\mathrm{p}]. \qquad \text{Equation 2}$$

In the following, both an isotropic as well as a cubic stiffness tensor will be used. In the case of an isotropic material, the stiffness is fully characterized by Young's modulus $E$ and Poisson's ratio $\nu$, while for the cubic material symmetry the constants are $C_{\{11\}}$, $C_{\{12\}}$ and $C_{\{44\}}$ in Voigt notation.

For the evolution of the internal variables, two plasticity models are used. The first is a phenomenological, rate-independent von Mises theory, commonly used for metallic materials to describe the behaviour of elasto-plastically deforming polycrystals [34]. The second one is a crystal-plasticity model [35]. The detailed material descriptions for these models and their application in FEM simulations is described later. First, the relevant equations are summarized. The von Mises model relies on the isotropic yield function

$$\varphi = \sqrt{\tfrac{3}{2}} \|\boldsymbol{\sigma}'\| - \sigma_\text{Y}(\varepsilon_\text{pe}) \leq 0, \qquad \text{Equation 3}$$

where $\|\boldsymbol{\sigma}'\|$ denotes the Frobenius norm of the deviatoric Cauchy stress tensor and $\sigma_\text{Y}(\varepsilon_\text{pe})$ is the yield strength depending on the accumulated plastic strain $\varepsilon_\text{pe}$. The plastic strains and the equivalent plastic strain develop according to an associated flow rule as [34]

$$\dot{\boldsymbol{\varepsilon}}_\text{p} = \gamma \frac{\boldsymbol{\sigma}'}{\|\boldsymbol{\sigma}'\|}; \quad \dot{\varepsilon}_\text{pe} = \sqrt{\tfrac{2}{3}}\, \gamma, \qquad \text{Equation 4}$$

where $\gamma \geq 0$ is the consistency parameter, that is only non-vanishing when the yield condition and the loading condition are fulfilled. The solution of these evolution equations is performed using the built-in material library of Abaqus in a time-implicit setting [32].

The equations of the crystal plasticity model (CP-FEM) are based on the same additive split of the strain tensor and Hooke's law. However, the phenomenological yield function and the flow rule are replaced by a viscoplastic flow rule in the slip systems accounting for the discrete slip systems and slip activity of a single crystal. A thorough presentation of details regarding CP-FEM is given in Ref. [36] and only briefly summarized here. The model relies on a rate-dependent theory [37]. The Chaboche flow rule is chosen [35]

$$\dot{\gamma}_\alpha = \dot{\gamma}_0\, \text{sgn}(\tau_\alpha) \langle \tfrac{|\tau_\alpha| - \tau_\text{sl}}{\tau_\text{D}} \rangle^m, \qquad \text{Equation 5}$$

where $\dot{\gamma}_\alpha$ is the slip rate in slip system $\alpha$, $\dot{\gamma}_0$ is a reference strain rate, $\tau_\text{D}$ is a constant drag stress, $m$ is the strain-rate sensitivity exponent, the resolved shear stress is $\tau_\alpha$ and the critical shear stress for dislocation slip is $\tau_\text{sl}$. The Maccauly bracket $\langle * \rangle$ implies $\max(*, 0)$. Consistent with the activation of octahedral slip systems in fcc metals and alloys, all slip systems experience an identical $\tau_\text{sl}$. No hardening is assumed with ongoing plastic deformation and thus, $\tau_\text{sl} = \text{const.}$. The resolved shear stress on each slip system is computed by

$$\tau_\alpha = \boldsymbol{\sigma} \cdot \boldsymbol{M}_\alpha, \qquad \text{Equation 6}$$

where $M_\alpha = d_\alpha \otimes_s n_\alpha$ is the Schmid tensor of slip system $\alpha$. The total plastic strain rate is

$$\dot{\boldsymbol{\varepsilon}}_\text{p} = \sum_\alpha \dot{\gamma}_\alpha\, \boldsymbol{M}_\alpha \qquad \text{Equation 7}$$

and the accumulated plastic slip rate is computed by

$$\dot{\gamma}_\text{acc} = \sum_\alpha |\dot{\gamma}_\alpha|. \qquad \text{Equation 8}$$

It is necessary to point out that by choosing an exponent $m > 15$, this rate-dependent model approaches a rate-independent model [36]. In this work, the exponent is chosen to be $m = 20$, ensuring that the CP-FEM simulations resemble a rate-independent setting. The CP material model is implemented using the User defined MATerial (UMAT) functionality of Abaqus in a time-implicit setting using an implicit Euler scheme.

A short comment regarding the FEM model with a schematic shown in Figure 4: The experimental set-up exists out of a cuboid sample and a spherical indenter. Within the FEM model, symmetry boundary conditions are applied within the SD-ND-plane, modelling half of the geometry and reducing computational effort. The sample is discretized using reduced integration linear hexahedral elements with a constant edge length of 5 μm, while the deformable indenter (considered to be linear elastic; SiC: $E$ = 420 GPa and $v$ = 0.15 [38]) is discretized using quadratic tetrahedral elements and the element edge length is refined to 1 μm at the contact area, in order to have a high accuracy there. Preliminary studies have shown that this mesh resolution is sufficient and no further deviation in the fields of interest occurs for finer meshes. The Abaqus built-in contact formulation of a hard surface-to-surface contact is chosen with the indenter as the master surface and the experimental friction coefficient is chosen to be $\mu = 0.2$, this being the experimental value. The sample is pinned at the bottom, while the load on the indenter is applied by means of a reference point and a tie constraint. Two simulation steps are used. In the first, the normal force is ramped up on the indenter to 2 N over the time span of 1 s. Subsequently, a displacement in $e_x$ of $u_x = 0.5$ mm is applied, which correspond to a sliding velocity of v=500µm/s for 1s. The location-dependent stress fields are evaluated directly on the symmetry plane, the SD-ND plane at the middle of the wear track.

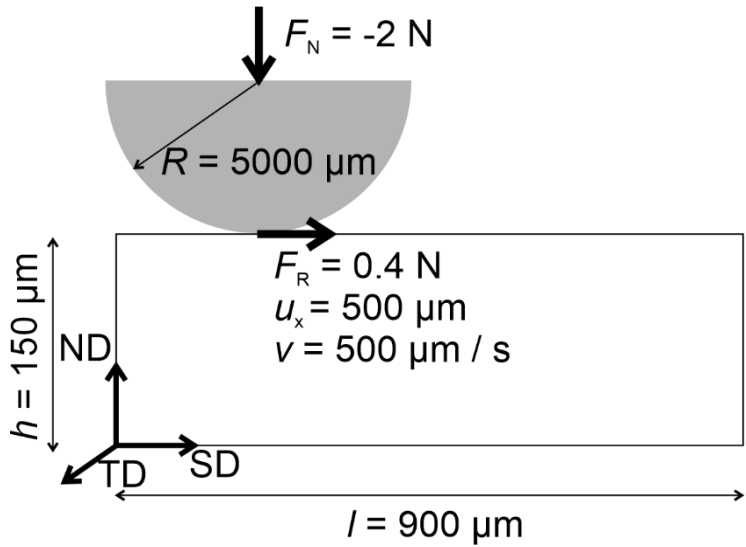

Figure 4. General FEM model of the tribological contact showing the notation of the coordination system and the used loading condition. Symmetric boundary conditions in the middle of the wear track are used within the SD-ND plane.

Prior to calculating the resolved shear stresses, all twin systems $(hkl)[uvw]$ (with a twin plane normal $[hkl]$) were defined in such a way that a positive resolved shear stress results in twin activation and a negative resolved shear stress in an AA-stacking, which is not possible form the materials perspective. The resolved shear stress $\tau$ on the twin systems were calculated using Equation 6. As the stress tensor is dependent on the lateral position, only the maxima of the resolved shear stress were extracted and investigated.

The material descriptions are presented in *Table 2*. The names of the various material models are a short representation of each model (LY = low yield strength, HY = high yield strength, noW = no work hardening behaviour, LW = low work hardening behaviour, HW = high work hardening behaviour, CP-FEM = CP-FEM was used considering elastic and plastic anisotropy). The stress strain curves of LYnoW, LYLW and LYHW are presented in *Figure 5* illustrating the differences within the work hardening behaviours. The elastic material properties vary only little in literature, whereby a larger spread is found for the parameters describing plasticity. The yield strength values are upper and lower limits in literature [27] (high yield strength measured with a grain size of 4 µm and the low yield strength corresponds to the resistance of the lattice to dislocation motion extracted from the Hall-Petch relationship). The critical resolved shear stress for dislocation slip $\tau_{sl}$ is the lowest found in literature [29]. Furthermore, $\tau_{sl}$ is connected by the Taylor factor to the low yield strength value which allows to investigate the effect of plastic anisotropy by comparing the material models LYnoW and CP. As the crystal orientation ND[$0\bar{1}\bar{1}$]/SD[$\bar{2}11$] is not symmetric with respect to the SD-ND-plane, no result exits with CP-FEM.

Table 2. Material model descriptions used in the finite element calculations. The abbreviations are used throughout the whole manuscript.

| Designation | **HYLW** | **LYHW** | **LYLW** | **LYnoW** | **CP** |
|---|---|---|---|---|---|
| **Elastic model** | isotropic | | | | anisotropic |
| | | $E$ = 203 GPa [28] $\nu$ = 0.25 [28] | | | $C_{\{11\}}$ = 196 GPa $C_{\{12\}}$ = 118 GPa $C_{\{44\}}$ = 130 GPa [39] |
| **Plastic model** | isotropic | | | | anisotropic |
| **Yield strength** | high | Low | | | low |
| | $\sigma_y$ = 360 MPa [27] | $\sigma_y$ = 125 MPa [27] | | | $\tau_{sl}$ = 43 MPa [29] |
| **Work hardening** | low | high | Low | none | none |
| | 2.5 GPa | 4.5 GPa | 2.5 GPa | 0 GPa | 0 GPa |

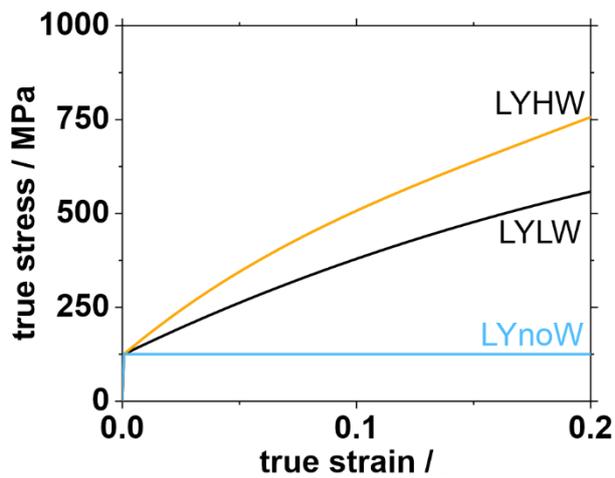

Figure 5. Strain-stress curves for the material behaviors labeled LYHW, LYLW and LYnoW. The work hardening behaviors of LYHW and LYLW were extracted from [28].

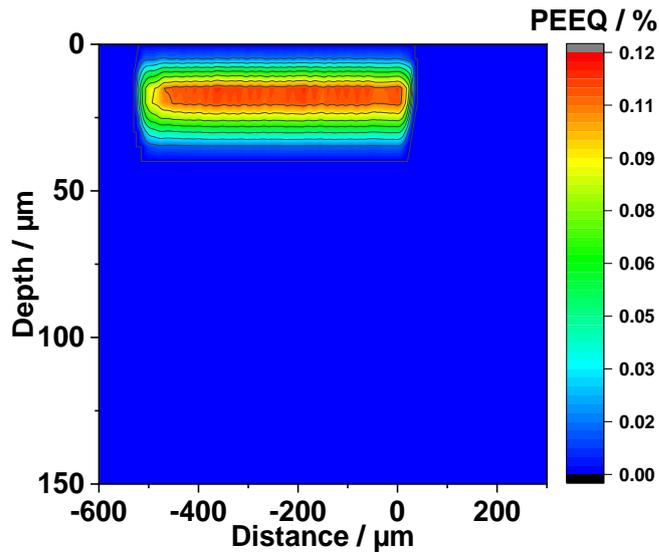

Figure 6. PEEQ calculated with HYLW; at the surface no strain is observed.

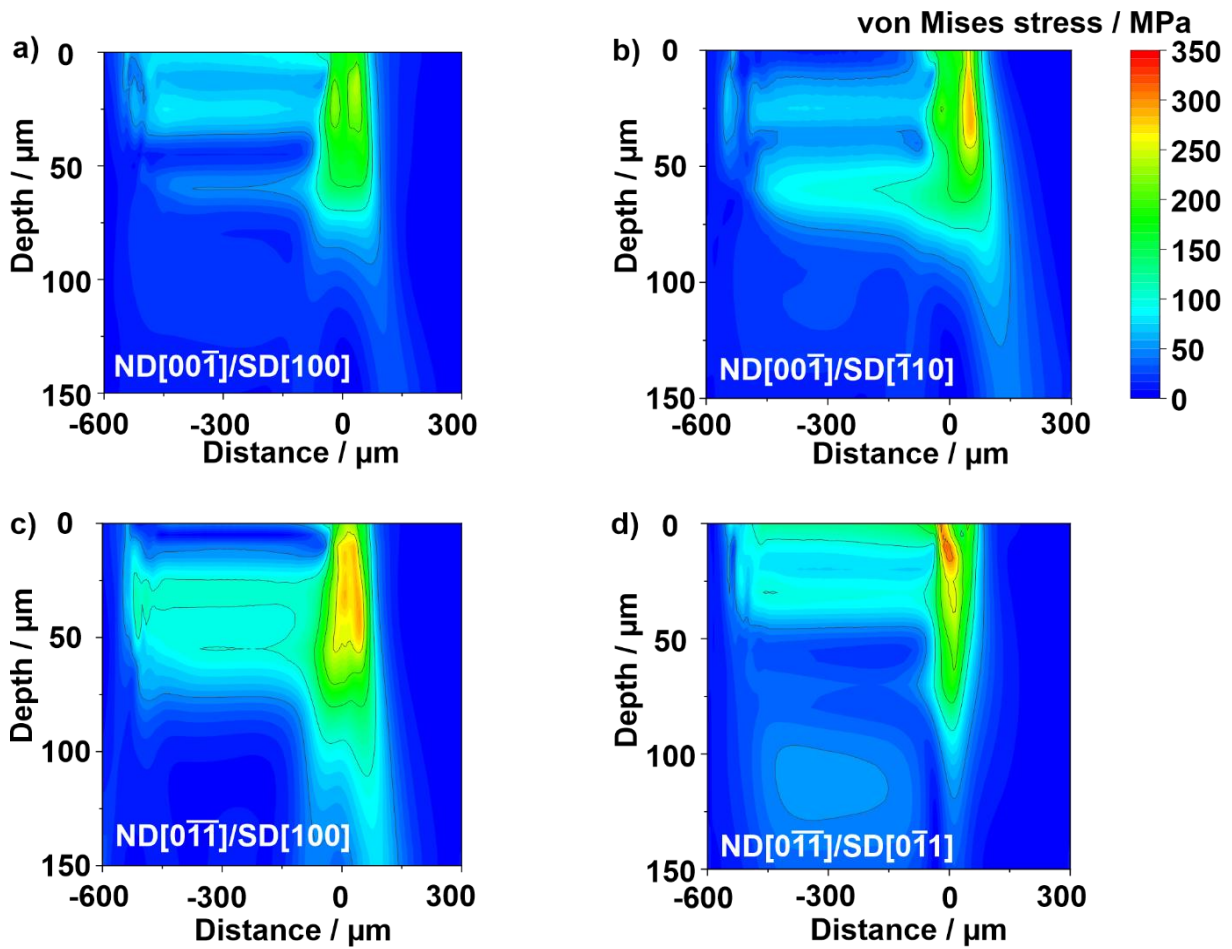

Figure 7. von Mises stresses calculated for the various crystallographic orientations using CP-FEM; a) ND[00$\bar{1}$]/SD[100], b) ND[00$\bar{1}$]/SD[$\bar{1}$10], c) ND[0$\bar{1}\bar{1}$]/SD[100] and d) ND[0$\bar{1}\bar{1}$]/SD[0$\bar{1}$1].